\documentclass[prc,twocolumn,showpacs,preprintnumbers,amsmath,amssymb,superscriptaddress,floatfix,nofootinbib]{revtex4}

\usepackage{graphicx}
\usepackage{amsmath}
\usepackage{amsfonts}
\usepackage{amssymb}
\usepackage{color}

\newcommand{\Slash}[1]{\ooalign{\hfil/\hfil\crcr$#1$}}

\begin{document}
\title{Nucleon pole contribution in the $pp\to ppK^+K^-$ reaction below the $\phi$ meson threshold }

\author{Qi-Fang L\"{u}} \affiliation{Department of Physics, Zhengzhou University, Zhengzhou, Henan 450001, China}

\author{Ju-Jun Xie}\email{xiejujun@impcas.ac.cn}
\affiliation{Institute of Modern Physics, Chinese Academy of
Sciences, Lanzhou 730000, China} \affiliation{Research Center for
Hadron and CSR Physics, Institute of Modern Physics of CAS and
Lanzhou University, Lanzhou 730000, China}

\affiliation{State Key Laboratory of Theoretical Physics, Institute
of Theoretical Physics, Chinese Academy of Sciences, Beijing 100190,
China}

\author{De-Min Li}\email{lidm@zzu.edu.cn} \affiliation{Department of Physics, Zhengzhou University, Zhengzhou, Henan 450001, China}

\begin{abstract}

Nucleon pole contribution in the $pp\to ppK^+K^-$ reaction below the
threshold of the production of the $\phi$ meson is studied within
the effective Lagrangian approach. It is assumed that the $K^- p$
final state originates from the decay of the hyperons
$\Lambda(1115)$ and $\Lambda(1405)$. In addition to the $pp$ final
state interaction (FSI) parametrized using the Jost function, we
have also considered the $K^+K^-$ FSI with the techniques of the
chiral unitary approach, where the scalar mesons $f_0(980)$ and
$a_0(980)$ were dynamically generated. Hence, the contributions from
scalar mesons $f_0(980)$ and $a_0(980)$ occur through the $K^+ K^-$
FSI. It is shown that the available experimental data are well
reproduced, especially the total cross sections and the invariant
mass distributions of $pp$ and $K^+K^-$. Furthermore, different
forms of the couplings (pseudoscalar and pseudovector) for the $\pi
NN$ interaction and different strengths for the proton-proton FSI
are also investigated. It is found that the contributions from
hyperon $\Lambda(1115)$ and $\Lambda(1405)$ are different between
these two kinds of couplings. On the other hand, the effects of the
proton-proton FSI can be adjusted by the cut off parameters used in
the form factors.

\end{abstract}
 \maketitle

\section{Introduction}{\label{introduction}}

The meson production reaction in nucleon-nucleon collisions near
threshold has the potential to yield information on hadron
properties~\cite{Hanhart:2003pg} and also plays an important role
for exploring the baryon spectroscopy~\cite{Zou:2009wp}. In recent
years, the experimental database on the reaction of $pp \to pp
K^+K^-$ near threshold has been expanded significantly. In addition
to the measurements of the $pp \to pp K^+K^-$ total and differential
cross sections, below the threshold of the production of the $\phi$
meson, performed experimentally with
COSY-11~\cite{Quentmeier:2001ec,Winter:2006vd} and
ANKE~\cite{Ye:2013ndq} detectors at the cooler synchrotron COSY in
Germany, there are invariant mass distributions of various
subsystems obtained at excess energies $\varepsilon = 10$, $23.9$,
and $28$ MeV~\cite{Ye:2013ndq,Silarski:2009yt} and in Dalitz
plots~\cite{Silarski:2013rfa,Silarski:2009yt}. The total and
differential cross sections are also available for the $pp \to
ppK^+K^-$ reaction above the $\phi$ meson threshold determined by
the ANKE~\cite{Maeda:2007cy,Ye:2012ae} Collaboration and the
DISTO~\cite{Balestra:1999yh} Collaboration.

In response to this wealth of data there have been theoretical
investigations for the $pp \to ppK^+K^-$ reaction above $\phi$ meson
production~\cite{Dzyuba:2008fi,Xie:2010md,Lebiedowicz:2011tp}.
However, the theoretical investigations of this reaction below the
$\phi$ meson threshold are scarce. Below the $\phi$ meson threshold,
the main contribution to the production of $K^+K^-$ pair could be
through the scalar mesons $a_0(980)$ and $f_0(980)$, thus, the
original motivation for the study of the $pp \to ppK^+K^-$ reaction
near threshold was to investigate the enigmatic properties of the
scalar resonances $a_0(980)$ and
$f_0(980)$~\cite{Silarski:2009yt,Ye:2013ndq}.

Unlike the production of the $\phi$ meson above threshold, in the
low energy region we do not need to separate the non$-\phi$ from the
$\phi$ contribution, and the fact that the data were spread over a
wide range of $K^+K^-$ invariant masses gives a special advantage to
investigation of the scalar mesons $a_0(980)$ and
$f_0(980)$~\cite{Ye:2013ndq}. These two mesons, which have been
studied by a large number of theoretical works, are commonly
explained as conventional $q \bar{q}$ mesons in the constituent
quark model~\cite{Morgan:1993td}, tetraquark states by
Jaffe~\cite{Jaffe:1976ig}, and $K \bar{K}$
molecules~\cite{Weinstein:1990gu}. Besides, within the chiral
unitary approach, the $f_0(980)$ and $a_0(980)$ scalar mesons are
dynamically generated from the interaction of $K \bar{K}$, $\pi
\pi$, and $\eta \pi$ treated as coupled channels in $I=0$ and $I=1$,
respectively~\cite{Oller:1997ti,Oller:1997ng,Oller:1998zr,GomezNicola:2001as,Pelaez:2006nj,Kaiser:1998fi}.
Both couple strongly to the $K \bar{K}$ channel. Inspired by those
results obtained from the chiral unitary approach, for the $pp \to
pp K^+ K^-$ reaction, we take the final state interaction (FSI)
between $K^+$ and $K^-$ into account by using the techniques of the
chiral unitary approach as in Refs.~\cite{Oller:1997ti,Li:2003zi}.
In this sense the contributions from scalar mesons $f_0(980)$ and
$a_0(980)$ occur through the $K^+ K^-$ FSI. This approach has been
used in the investigation of the FSI of mesons in different
processes in order to get a better understanding of the nature of
the meson resonances as shown in
Refs.~\cite{Marco:1999df,Markushin:2000fa,Palomar:2003rb,Guo:2006ai}

It has been suggested that the $\Lambda(1405)$ could play an
essential role on the kaon pair production through the $pp \to
pK^+(\Lambda(1405) \to K^- p)$ process~\cite{Wilkin:2008yp} and this
process seems more important than the contributions from the scalar
mesons~\cite{Dzyuba:2008fi,Bratkovskaya:1998mj}. Indeed, the role
played by the $\Lambda(1405)$ state is crucial for reproducing the
$K^-p$ mass distribution~\cite{Geng:2007vm,Xie:2010md}. In
Ref.~\cite{Xie:2010md}, the reaction of $pp \to ppK^+K^-$ has been
studied by assuming that the $K^-p$ final state originates from the
decay of the $\Lambda(1405)$, where the $N_{1/2^-}^*(1535)$
resonance acts as a doorway state for the production of
$\Lambda(1405)$. However, the model calculations of
Ref.~\cite{Xie:2010md} underestimate the total cross sections of the
$pp \to ppK^+K^-$ reaction near the kinematical threshold (see Fig.
4 of Ref.~\cite{Xie:2010md}). So, in the present work, within the
effective Lagrangian approach, we restudy the $pp \to ppK^+K^-$
reaction below the threshold of the $\phi$ meson production by
considering the contribution from the nucleon pole. Additionally,
different forms of the couplings (pseudoscalar and pseudovector) for
the $\pi NN$ interaction and different strength for the
proton-proton FSI are also investigated.

In the next section, we will present the formalism and ingredients
necessary for our estimations, then numerical results and
discussions are given in Sec. III. Finally, a short summary is given
in the last section.

\section{Formalism and ingredients}{\label{formalism}}

\begin{figure*}[htbp]
\begin{center}
\includegraphics[scale=0.6]{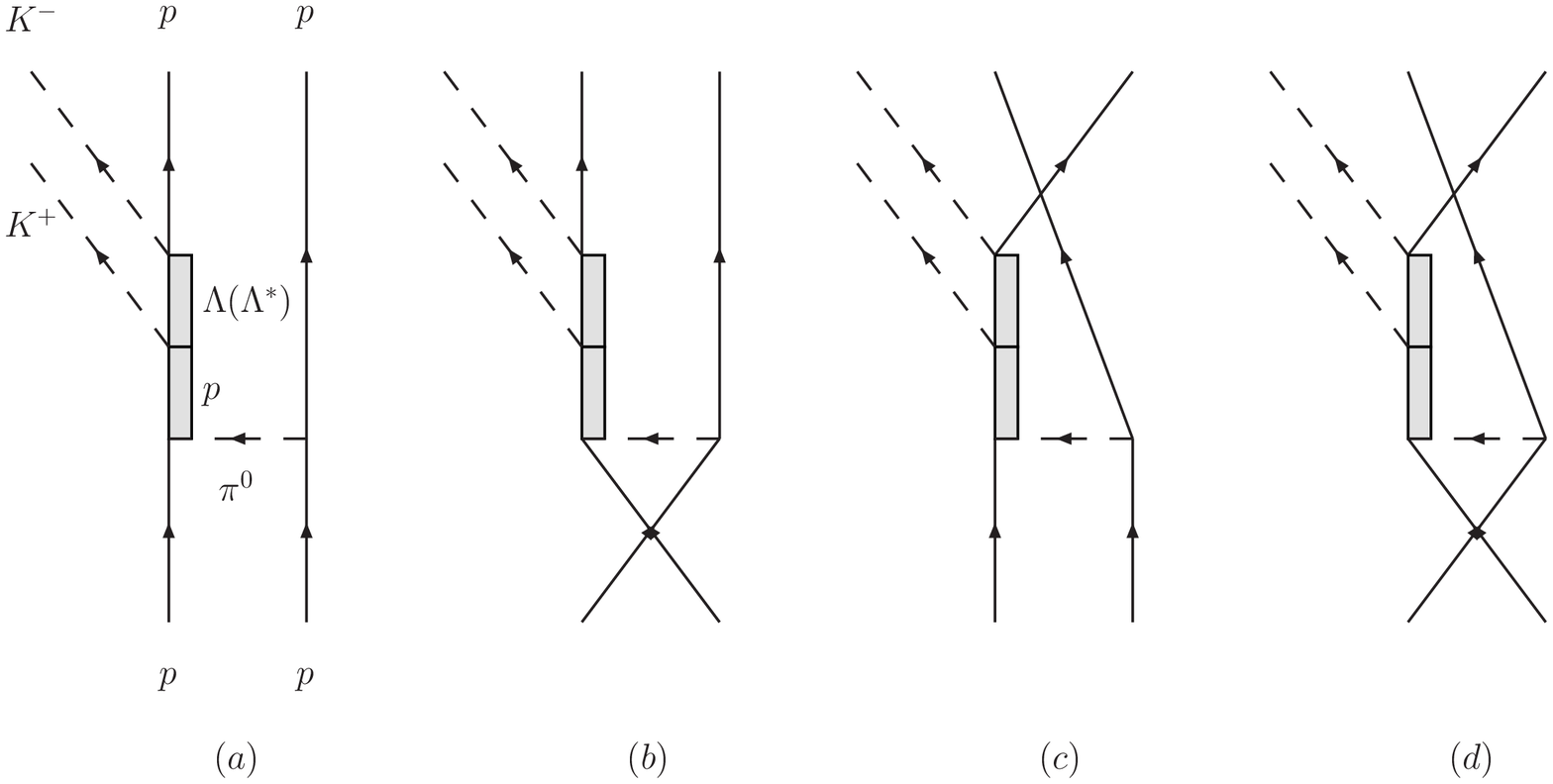}
\vspace{-0.2cm} \caption{Feynman diagrams for the $pp \to ppK^+K^-$
reaction.} \label{diagram}
\end{center}
\end{figure*}

We study the $pp \to ppK^+K^-$ reaction below the threshold of the
production of the $\phi$ meson within an effective Lagrangian
approach. The basic Feynman diagrams for this process are depicted
in Fig.~\ref{diagram}, where we pay attention to the contribution
from the nucleon pole for the production of the $K\Lambda(1115)$
pair and $K\Lambda(1405)$ pair, while the $K^-p$ pair is produced by
the decay of the off-shell $\Lambda(1115)$ and the subthreshold
$\Lambda(1405)$ ($\equiv \Lambda^*$). Because of the large $\pi NN$
coupling and the small pion mass, the underlying mechanism will be
dominated by the $\pi^0$ exchange. Thus, the contributions from the
$\eta$, $\rho$, and $\omega$ exchanges are neglected in the present
calculation.

To compute the amplitudes of these diagrams shown in
Fig.~\ref{diagram}, we need the effective Lagrangian densities for
the interaction vertexes. There are two forms for $\pi NN$
interaction commonly employed in the literature~\cite{Liang:2004sd}.
One is the pseudoscalar (PS) coupling,
\begin{equation}
{\cal L}^{PS}_{\pi N N}  = -i g_{\pi N N} \bar{\psi}_N \gamma_5 \vec\tau
\cdot \vec\pi \psi_N, \label{pinps}
\end{equation}
and the other one is the pseudovector (PV) coupling,
\begin{equation}
{\cal L}^{PV}_{\pi N N}  = - \frac{g_{\pi N N}}{2m_N} \bar{\psi}_N \gamma_5 \gamma_{\mu} \vec\tau
\cdot \partial^{\mu} \vec\pi \psi_N. \label{pinpv}
\end{equation}

Following the SU(3) flavor symmetry, the $KN\Lambda$ interaction
Lagrangian densities are similar to the $\pi NN$ interaction,
\begin{equation}
{\cal L}^{PS}_{K N \Lambda}  = -i g_{K N \Lambda} \bar{\psi}_N
\gamma_5 K \psi_{\Lambda} + {\rm h.c.}, \label{knps}
\end{equation}
\begin{equation}
{\cal L}^{PV}_{K N \Lambda}  = - \frac{g_{K N
\Lambda}}{m_N+m_{\Lambda}} \bar{\psi}_N \gamma_5 \gamma_{\mu}
\partial^{\mu}K \psi_{\Lambda} + {\rm h.c.}. \label{knpv}
\end{equation}

In addition, the effective $KN\Lambda(1405)$ coupling is also
needed~\cite{Xie:2013wfa},
\begin{equation}
{\cal L}_{K N \Lambda^*}  = -i g_{K N \Lambda^*} \bar{\psi}_N K
\psi_{\Lambda^*}+ {\rm h.c.}. \label{knlams}
\end{equation}

The coupling constants in the above Lagrangian densities are taken
as~\cite{Xie:2013wfa}: $g_{\pi NN} = 13.45$, $g_{K N \Lambda} =
-13.98$, and $g_{K N \Lambda^*} = 1.51$.

On the other hand, we need to include the form factors because the
hadrons are not point-like particles. We adopt here the common
scheme used in many previous works. In our calculation, for the $\pi
NN$ vertex, we take the form following that used in
Refs.~\cite{Machleidt:1987hj,Machleidt:1989tm,Brockmann:1990cn},
\begin{equation}
F_{\pi}(k^2_{\pi})=\frac{\Lambda^2_{\pi}-m_{\pi}^2}{\Lambda^2_{\pi}- k_{\pi}^2},
\end{equation}
where $k_{\pi}$, $m_{\pi}$, and $\Lambda_{\pi}$ are the
four-momentum, mass, and cut-off parameter for the exchanged pion
meson, respectively. For the cut-off parameter $\Lambda_{\pi}$, we
take the value of $1.3$ GeV~\cite{Xie:2007vs,Xie:2007qt}.

The form factors for the off-shell nucleon, and the hyperon
$\Lambda(1115)$ and $\Lambda(1405)$ states are taken in the form
advocated in
Refs.~\cite{Feuster:1997pq,Penner:2002ma,Shklyar:2005xg},
\begin{equation}
F(q^{2}_{ex},M_{ex}) = \frac{\Lambda^4}{\Lambda
^4+(q^{2}_{ex}-M^2_{ex})^2},
\end{equation}
where $q_{ex}$ and $M_{ex}$ are the four-momentum and the mass of
the exchanged hadron, respectively. In the present calculation, to
minimize the number of free parameters, we use the same cut-off
parameters for those hadrons for simplicity, ${\it i.e.}$,
$\Lambda_N = \Lambda_{\Lambda} = \Lambda_{\Lambda^*} = \Lambda$. The
value of the cut off parameter will be discussed in the following.

Then, according to the Feynman rules, the scattering amplitudes for
the $pp \to pp K^+ K^-$ reaction can be obtained straightforwardly
with the above effective couplings. Here, we give explicitly the
amplitude of Fig.~\ref{diagram} (a) with $\Lambda(1115)$ exchange
and in the case of PS coupling for $\pi NN$ and $KN\Lambda(1115)$
vertexes as an example,
\begin{eqnarray}
{\cal M}^{\Lambda}_{\rm a} & = & g_{\pi NN}^2 g_{K N \Lambda}^2
F_{\pi}^2(k^2_{\pi}) F(q_1^2,m_N) F(q_2^2,m_{\Lambda})
G_{\pi}(k_{\pi}) \nonumber\\ &&  \times \bar{u} (p_4,s_4) \gamma_5
G_{\Lambda}(q_{\Lambda}) \gamma_5 G_N(q_N) \gamma_5 u(p_1,s_1) \nonumber\\
&&  \times \bar{u}(p_3,s_3) \gamma_5 u(p_2,s_2) ,
\end{eqnarray}
where $s_i~(i=1,2,3,4)$ and $p_i~(i=1,2,3,4)$ represent respectively
the spin projection and four-momentum of the initial or final
protons; $G_{\pi}(k_{\pi})$ [$G_N(q_N)$] is the propagator for the
exchanged $\pi$ meson [nucleon].

The $\pi$ meson propagator used in our calculation is
\begin{equation}
G_{\pi}(k_{\pi})=\frac{i}{k_{\pi}^2-m^2_{\pi}},
\end{equation}

The propagators of the nucleon and $\Lambda(1115)$ can be written as
\begin{equation}
G_{N/\Lambda}(q_{N/\Lambda}) = \frac{i(\Slash q_{N/\Lambda} +
m_{N/\Lambda})}{q_{N/\Lambda}^2 - m_{N/\Lambda}^2},
\end{equation}
where the  $q_N$ [$q_{\Lambda}$]is the four-momentum of the
intermediate nucleon [$\Lambda(1115)$].

In addition, the propagator of the $\Lambda(1405)$ resonance is
written in a Breit-Wigner form~\cite{Liang:2002tk},
\begin{equation}
G_{\Lambda^*}(q)=\frac{i(\Slash q +
m_{\Lambda^*})}{q^2-m^2_{\Lambda^*}+im_{\Lambda^*}\Gamma_{\Lambda^*}(q^2)},\label{G1405}
\end{equation}
where $\Gamma_{\Lambda^*}(q^2)$ is the energy-dependent total width
of the $\Lambda^*$ resonance, which can be expressed
as~\cite{Xie:2013wfa}
\begin{eqnarray}
&& \Gamma_{\Lambda^*} (q^2) = \frac{3g^2_{\Lambda^* \pi \Sigma}}{4
\pi}(E_{\Sigma}+m_{\Sigma}) \frac{|\vec p_{\Sigma}|}{\sqrt{q^2}} +
\nonumber \\
&& \frac{g^2_{\Lambda^* \bar{K} N}}{2 \pi}(E_N+m_N) \frac{|\vec
p_N|}{\sqrt{q^2}} \theta (\sqrt{q^2} - m_{\bar{K}} - m_N),
\end{eqnarray}
with
\begin{equation}
E_{\Sigma / N} = \frac{q^2 + m^2_{\Sigma / N}-m^2_{\pi /
\bar{K}}}{2\sqrt{q^2}},
\end{equation}
\begin{equation}
|\vec p_{\Sigma / N}| = \sqrt{E^2_{\Sigma / N}-m^2_{\Sigma / N}}.
\end{equation}

According to Fig.~\ref{diagram}, the full invariant amplitude for
the $pp \to ppK^+K^-$ reaction through the proton and
$\Lambda(1115)$ [ proton and $\Lambda(1405)$ ] is composed of four
parts:
\begin{eqnarray}
{\cal M}_0 = \sum_{i = a,~ b,~ c,~ d} \eta_i {\cal M}_i,
\end{eqnarray}
with the factors $\eta_a = \eta_d = 1$ and $\eta_b = \eta_c = -1$.

The final state interaction for the final $K^+K^-$ subsystem is
given by the meson-meson amplitude from the lowest order chiral
Lagrangian with the chiral unitary approach as in
Ref.~\cite{Oller:1997ti}. We choose five channels $K^+K^-$, $K^0
\bar{K}^0$, $\pi^+ \pi^-$, $\pi^0 \pi^0$ and $\pi^0 \eta$ which are
denoted from 1 to 5, to calculate the amplitude $T_{K^+K^- \to
K^+K^-}$ in the charge eigenstates directly.

The scattering amplitude $T_{K^+K^- \to K^+K^-}$ can be obtained by
solving the Bathe-Salpeter equation,~\footnote{As shown in
Ref.~\cite{Oller:1997ti}, the scattering amplitude $T_{K^+ K^- \to
K^+ K^-}$ is projected to be $S$-wave.}
\begin{eqnarray}
T = [1-VG]^{-1}V, \label{bs}
\end{eqnarray}
where $G$ is a diagonal matrix with the matrix elements
\begin{eqnarray}
G_{ii}= i\int \frac{d^4 q}{(2 \pi)^4} \frac{1}{q^2-m^2_{i_1}+i\epsilon} \frac{1}{(P-q)^2-m^2_{i_2}+i\epsilon} \nonumber\\
 = \int_{0}^{q_{max}} \frac{q^2 dq}{(2 \pi)^2} \frac{\omega_1+\omega_2}{\omega_1 \omega_2 [P^{02}-(\omega_1+\omega_2)^2+i\epsilon]},
\end{eqnarray}
where $P$ is the total four-momentum of the meson-meson system and
$q$ is the four-momentum of one of the intermediate mesons with
$\omega_i = ({\vec q}^2 + m_i^2)^{1/2}$. The loop integration
variable is regularized with a cutoff $|\vec q|<q_{max}$ and
$q_{max}$ = 1030 MeV as used in Refs.~\cite{Oller:1997ti,Li:2003zi}.
With this value, the scalar mesons $f_0(980)$ and $a_0(980)$ were
dynamically generated as poles of the $S$-wave amplitudes. Thus, in
the present case, the contributions from scalar mesons $f_0(980)$
and $a_0(980)$ occur via the FSI between $K^+$ and $K^-$.

For the FSI of the proton and proton in the final state, we use the
general framework based on the Jost function formalism,
\begin{eqnarray}
J(k)^{-1} = \frac{k+ i \beta}{k- i \alpha}, \label{fsipp}
\end{eqnarray}
where $k$ is the internal momentum of $pp$ subsystem. In this case,
we use two sets of parameters. One is the widely used $^1S_0$ $pp$
interaction, with $\alpha = -20.5$ MeV and $\beta = 166.7$
MeV~\cite{Sibirtsev:1999ka,Sibirtsev:2005zc}. The other is $\alpha =
0.1$ fm$^{-1}$ and $\beta = 0.5 $ fm$^{-1}$ (corresponding to
$\alpha = 19.7$ MeV and $\beta = 98.7$ MeV) as used in
Ref.~\cite{Maeda:2007cy}.

Taking the FSI of $K^+K^-$ and $pp$ subsystems into account, the
amplitude of the $pp\to pp K^+K^-$ reaction can be written
as,~\footnote{It is worth mentioning that the loop function
$G_{K^+K^-}$ and the amplitude $T_{K^+K^- \to K^+K^-}$ only depend
on the invariant mass of the $K^+K^-$ subsystem.}
\begin{eqnarray}
{\cal M} = \left( {\cal M}_0 + {\cal M}_0 G_{K^+K^-} T_{K^+K^- \to
K^+K^-} \right )J(k)^{-1}.
\end{eqnarray}

Then the calculations of the invariant scattering amplitude $|{\cal
M}|^2$ and the cross sections for $pp \to pp K^+ K^-$ reaction are
straightforward,
\begin{eqnarray}
&& d\sigma (pp\to ppK^+K^-) = \frac{1}{4}\frac{m^2_p}{F}
\sum_{s_1,s_2} \sum_{s_3,s_4} |{\cal M}|^2 \nonumber\\ && \times
\frac{m_p d^{3} p_{3}}{E_{3}} \frac{m_p d^{3} p_4}{E_4} \frac{d^{3}
p_{K^+}}{2 E_{K^+}} \frac{d^{3} p_{K^-}}{2 E_{K^-}} \nonumber\\ &&
\times \frac{1}{2}\delta^4(p_{1}+p_{2}-p_{3}-p_{4}-p_{K^+}-p_{K^-}),
\label{eqcs}
\end{eqnarray}
where $E_3$ and $E_4$ are the energies of the final protons;
$p_{K^+}$ and $E_{K^+}$ [$p_{K^-}$ and $E_{K^-}$] stand for the
four-momentum and energy of the final state $K^+$ [
$K^-$],respectively. The factor $\frac{1}{2}$ before the $\delta$
function comes from the two identical protons in the final state,
while the flux factor $F$ in the above equation is
\begin{eqnarray}
F=(2 \pi)^8\sqrt{(p_1 p_2)^2-m^4_p}~. \label{eqff}
\end{eqnarray}

Since the relative phase between $\Lambda(1115)$ and $\Lambda(1405)$
exchanges is not known, the interference terms between these parts
could be small and are ignored in our concrete calculation.

\section{Numerical results and discussions}

\begin{figure}[htbp]
\begin{center}
\includegraphics[scale=1.]{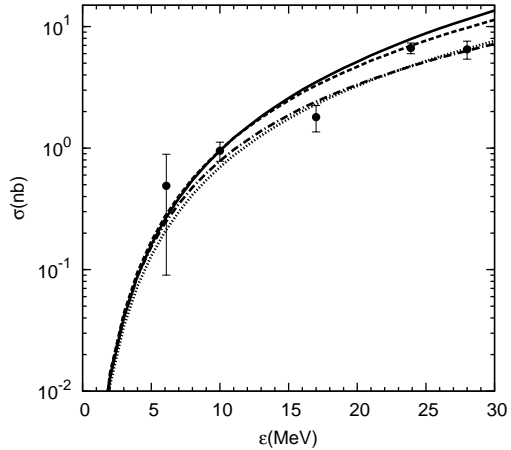}
\caption{Total cross sections vs excess energies ($\varepsilon$) for
the $pp \to ppK^+K^-$ reaction from the present calculation. The
experimental data are taken from
Refs.~\cite{Quentmeier:2001ec,Winter:2006vd,Ye:2013ndq}. The solid,
dashed, dotted, and dot-dashed curves stand for the results obtained
with the parameters of Sets I, II, III, and IV, respectively.}
\label{tcsa}
\end{center}
\end{figure}

With the formalism and ingredients given above, the total cross
section versus excess energy $\varepsilon$ for the $pp \to ppK^+K^-$
reaction is calculated by using a Monte Carlo multi-particle phase
space integration program. The results for $\varepsilon$ from the
reaction threshold to $30$ MeV, which is just below the $\phi$
threshold ($\varepsilon = 32$ MeV), and the experimental data taken
from Refs.~\cite{Quentmeier:2001ec,Winter:2006vd,Ye:2013ndq}, are
shown in Fig.~\ref{tcsa}. In our calculation, we take two types (PS
and PV) of $\pi NN$ and $KN\Lambda$ couplings, and two sets of
proton-proton FSI parameters. Therefore, there are a total of four
combinations as shown in Tab.~\ref{tab1}.

\begin{table}[htbp]
\begin{center}
\caption{ \label{tab1} Parameters used in the present calculation.}
\footnotesize
\begin{tabular}{|cccc|}
\hline
Set &  $\pi NN$ and  $KN \Lambda$ & $pp$ FSI  & Cut off\\
 &  couplings  & (MeV)  & (GeV)\\
\hline
I & PS &$\alpha = 19.7$, $\beta = 98.7$ & 1.5\\
II &PS &$\alpha = -20.5$, $\beta = 166.7$ & 1.3\\
III &PV &$\alpha = 19.7$, $\beta = 98.7$ & 1.5\\
IV  & PV &$\alpha = -20.5$, $\beta = 166.7$ & 1.3\\
\hline
\end{tabular}
\end{center}
\end{table}

In Fig.~\ref{tcsa}, the solid, dashed, dotted ,and dot-dashed curves
stand for our theoretical results obtained with the parameters of
Sets I, II, III, and IV, respectively. Because of the large error
bars of the experimental data points, from Fig.~\ref{tcsa} one can
see that with the cut-off parameters of form factors for exchanged
hadrons in the different sets listed in Table~\ref{tab1}, we can
reproduce the experimental data on the total cross sections of the
$pp \to pp K^+K^-$ reaction. Also, one can see that although the
absolute values of those parameters of different sets have some
discrepancies, they all can fairly well describe the experimental
data, but, the trend of the results obtained with PS and PV coupling
are different.

\begin{figure*}[htbp]
\begin{center}
\includegraphics[scale=1.1]{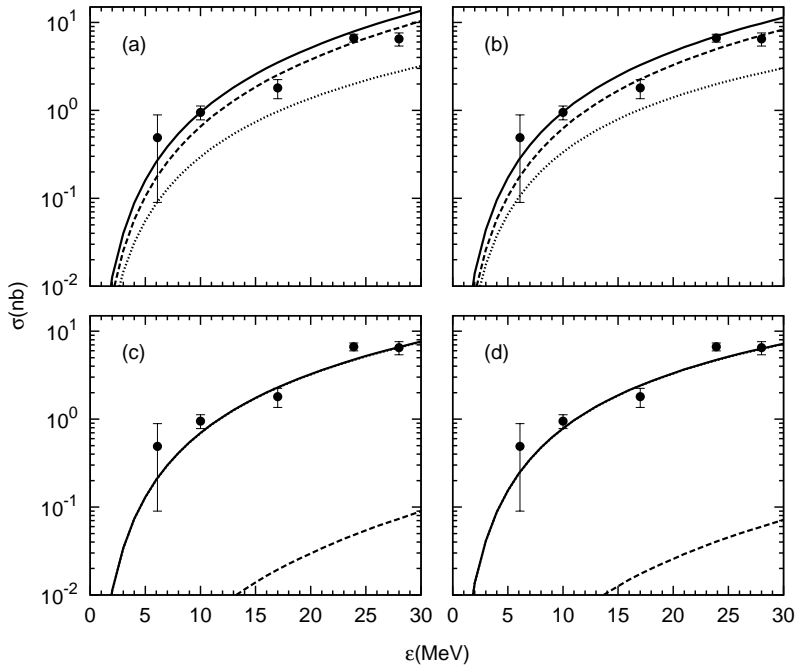}
\caption{Total cross sections for the $pp \to ppK^+K^-$ reaction.
The experimental data are taken from
Refs.~\cite{Quentmeier:2001ec,Winter:2006vd,Ye:2013ndq}. The dashed
and dotted lines stand for contributions from $\Lambda(1115)$ and
$\Lambda(1405)$, respectively. The individual panels are (a) results
obtained with Set I, (b) results obtained with Set II, (c) results
obtained with  Set III, and (d) results obtained with Set IV.}
\label{tcsb}
\end{center}
\end{figure*}

However, the contributions of $\Lambda(1115)$ and $\Lambda(1405)$
are different between PS and PV couplings. These results are
depicted in Fig.~\ref{tcsb}, where the dashed and dotted lines stand
for contributions from $\Lambda(1115)$ and $\Lambda(1405)$,
respectively. The results shown in individual panels (a), (b), (c)
,and (d) are obtained with Set I, II, III, and IV. It is shown that
the $\Lambda(1115)$ hyperon plays a dominant role in the case of PS
coupling, while the $\Lambda(1405)$ also has a significant
contribution. In contrast, in the case of PV coupling, the
$\Lambda(1405)$ contribution is predominant while the
$\Lambda(1115)$ hyperon contribution is rather small and can be
neglected.

Since we only pay attention to the $pp \to ppK^+K^-$ reaction below
the threshold of the production of the $\phi$ meson and also near
the $\Lambda(1405)$ threshold, it is expected that $\Lambda(1405)$
would play an important role in this energy
region~\cite{Maeda:2007cy,Xie:2010md,Geng:2007vm}. Although the PS
coupling can also reproduce the total cross section data, it seems
that the PV coupling is more favored. As shown in
Ref.~\cite{Liang:2004sd}, the PV coupling is more general than the
PS coupling. Besides, it is also shown that, for the $pp \to pp K^+
K^-$ reaction, the effect of the $pp$ FSI on the toal cross section
can be adjusted by modifying the cut off parameters in the form
factors of the intermediate proton and $\Lambda(1115)$.

\begin{figure*}[htbp]
\begin{center}
\includegraphics[scale=0.35]{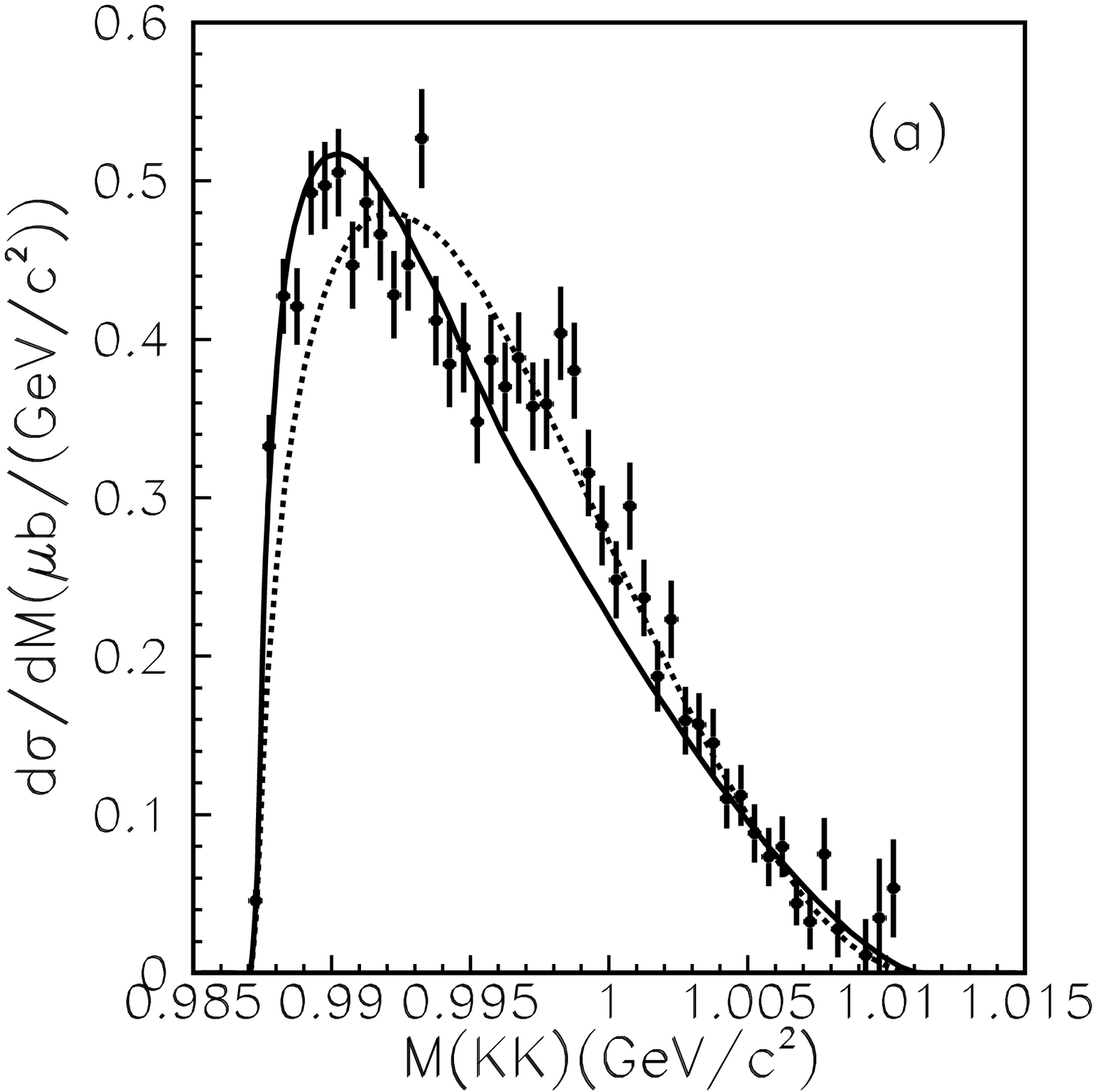}
\includegraphics[scale=0.35]{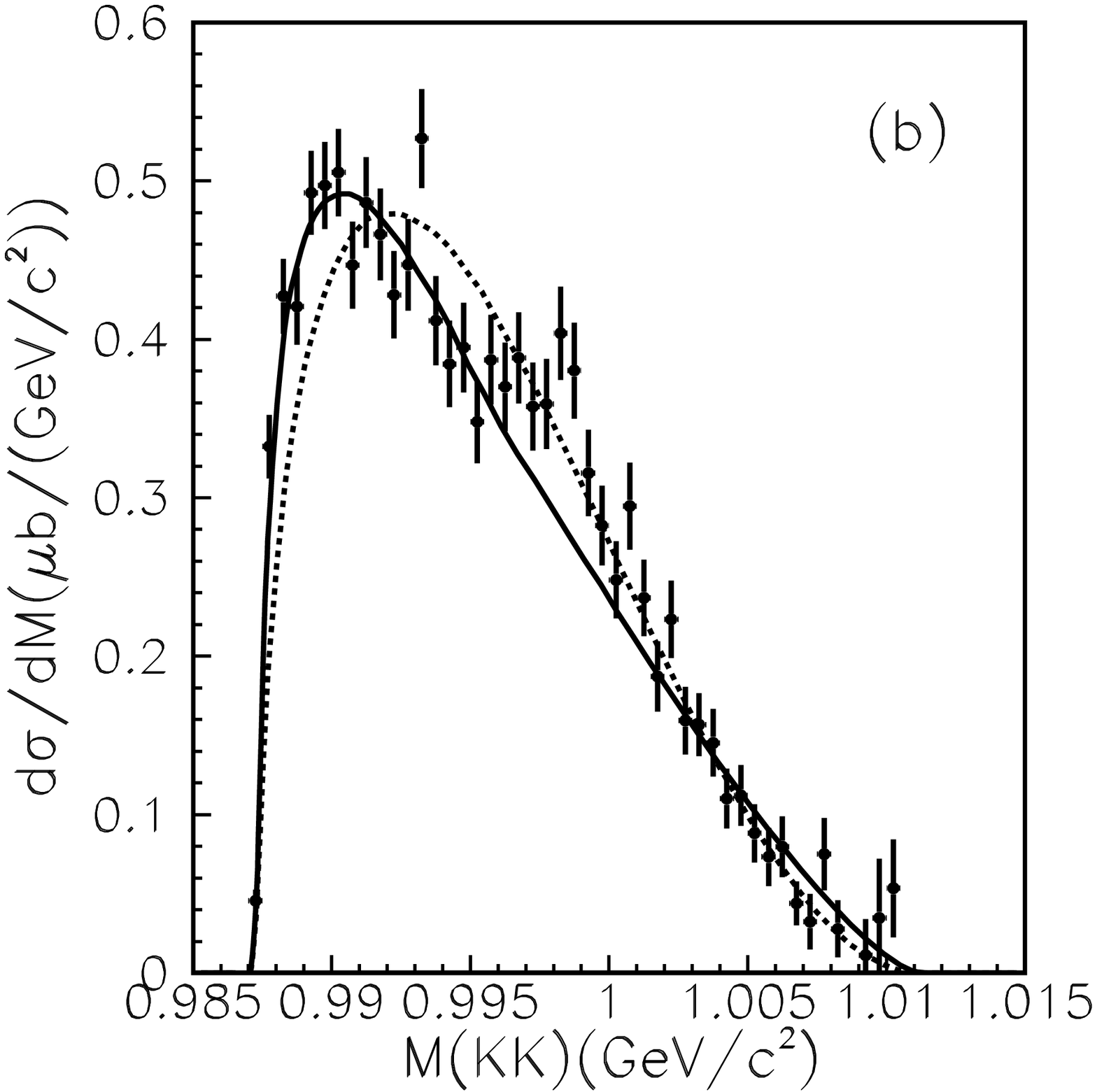}
\includegraphics[scale=0.35]{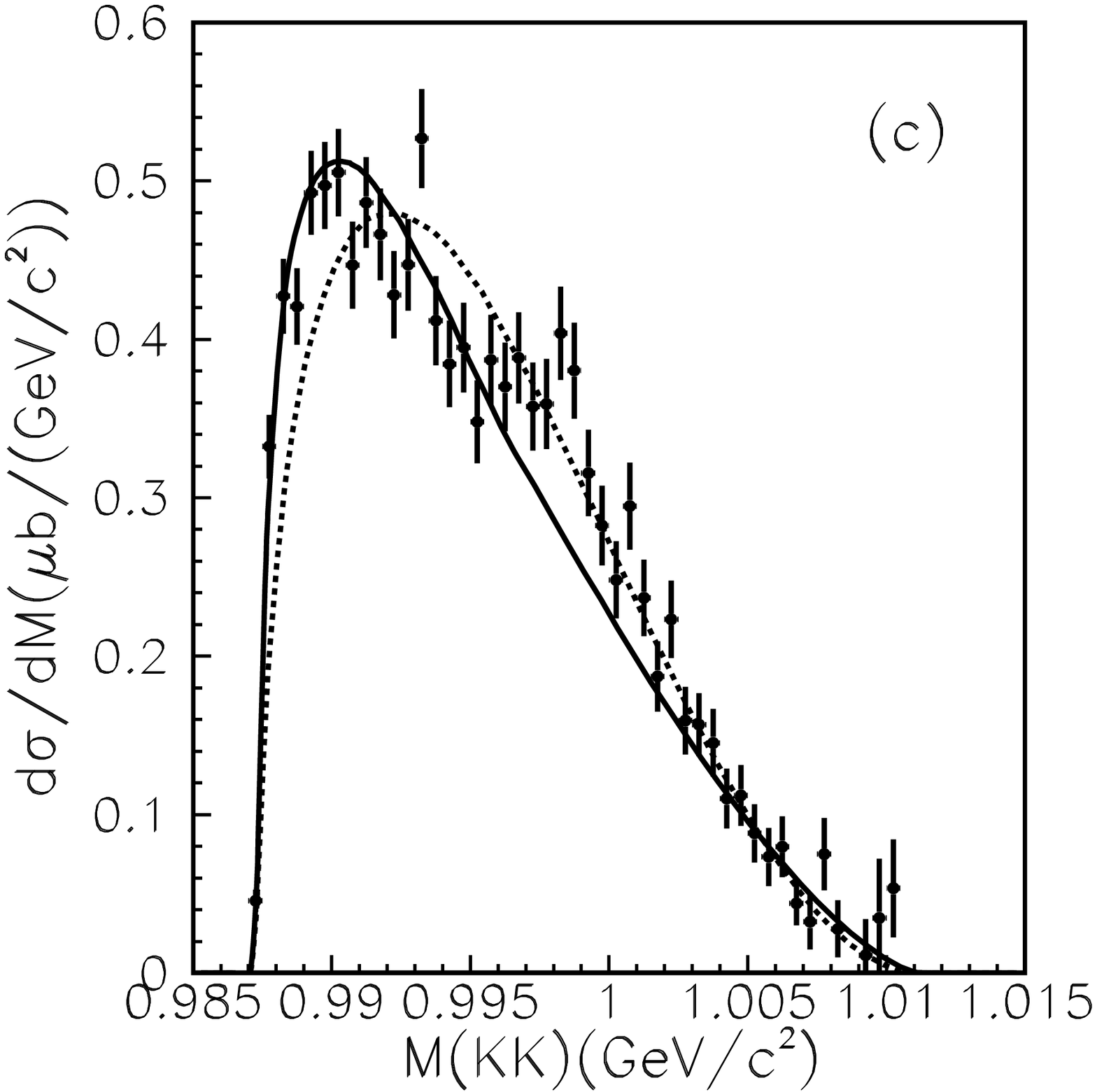}
\includegraphics[scale=0.35]{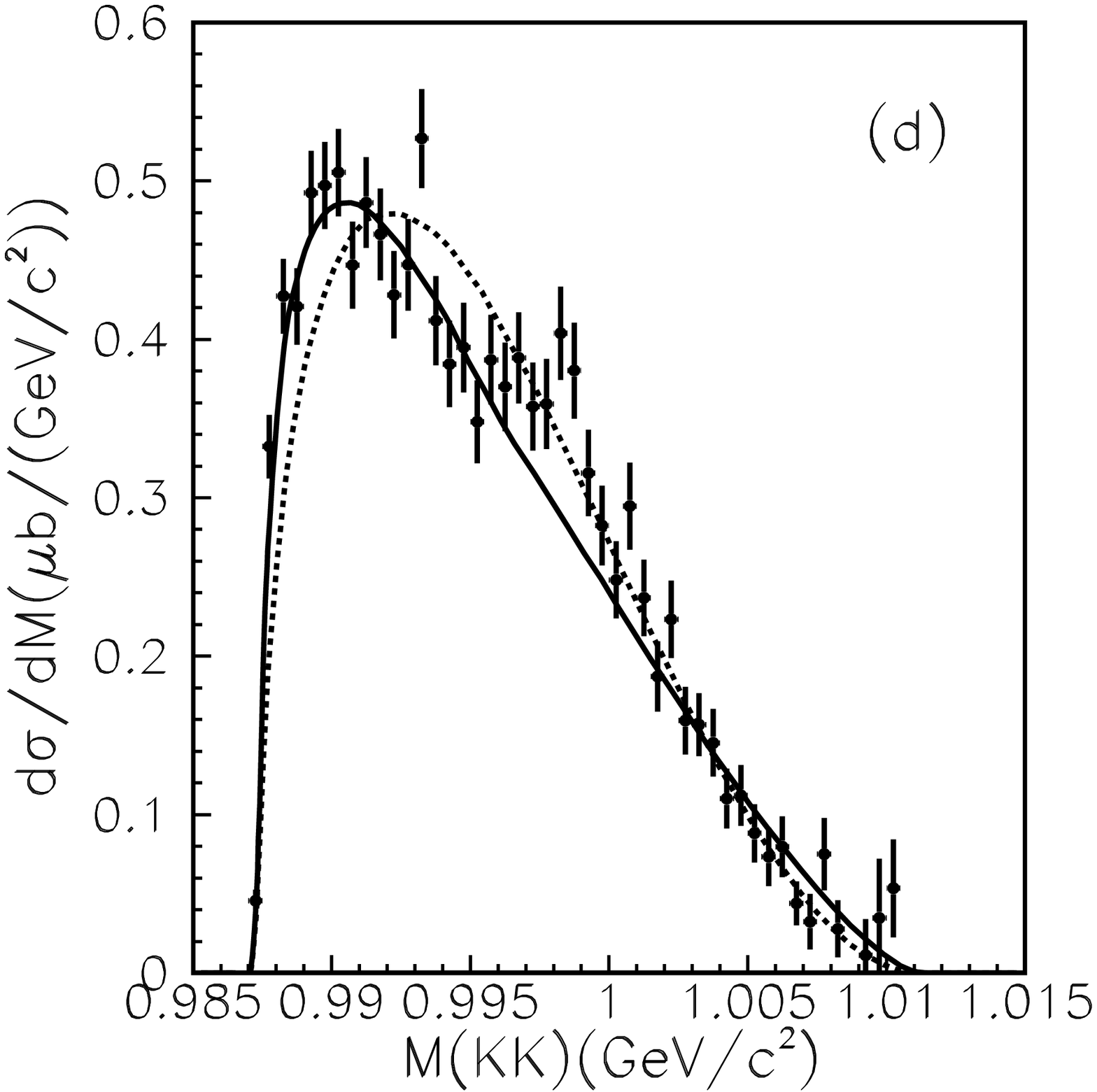}
\vspace{-0.2cm} \caption{The $K^+K^-$ invariant mass distribution
(solid lines) at the excess energy of $\varepsilon$ = 23.9 MeV
compared with the experimental data~\cite{Ye:2013ndq} and phase
space distribution (dashed lines). The panels (a), (b), (c), and (d)
denote the results obtained from Set I, II, III, and IV,
respectively. }\label{kk239}
\end{center}
\end{figure*}

To show the effect from the $K^+K^-$ FSI, we give the results for
the $K^+ K^-$ invariant mass spectrum of the $pp \to ppK^+K^-$
reaction at an excess energy $\varepsilon$ = 23.9 MeV in
Fig.~\ref{kk239}, where panels (a), (b), (c), and (d) stand for the
results obtained with the parameters of Sets I, II, III and, IV,
respectively. In Fig.~\ref{kk239}, the dashed lines are pure phase
space distributions, while the solid lines are full calculations
from our model. By comparing our theoretical results with the
experimental data, we found that the $K^+K^-$ FSI employed within a
chiral unitary approach plays an essential role in describing the
line shape of the $K^+K^-$. The peak near the $K^+K^-$ threshold can
be well reproduced by including the $K^+K^-$ FSI with the techniques
of the chiral unitary approach, where the scalar mesons $f_0(980)$
and $a_0(980)$ were dynamically generated. In this sense, the
$f_0(980)$ and $a_0(980)$ mesons play an important role in the $pp
\to pp K^+K^-$ reaction below the threshold of the production of the
$\phi$ meson. Furthermore, the $pp$ FSI can also slightly influence
the $K^+K^-$ invariant mass distribution. Here, we find again that
the PS and PV couplings are both good enough to reproduce the
experimental data, and the effects of $pp$ FSI on the differential
cross sections can also be adjusted by the cut off parameters.

\begin{figure*}[htbp]
\begin{center}
\includegraphics[scale=0.8]{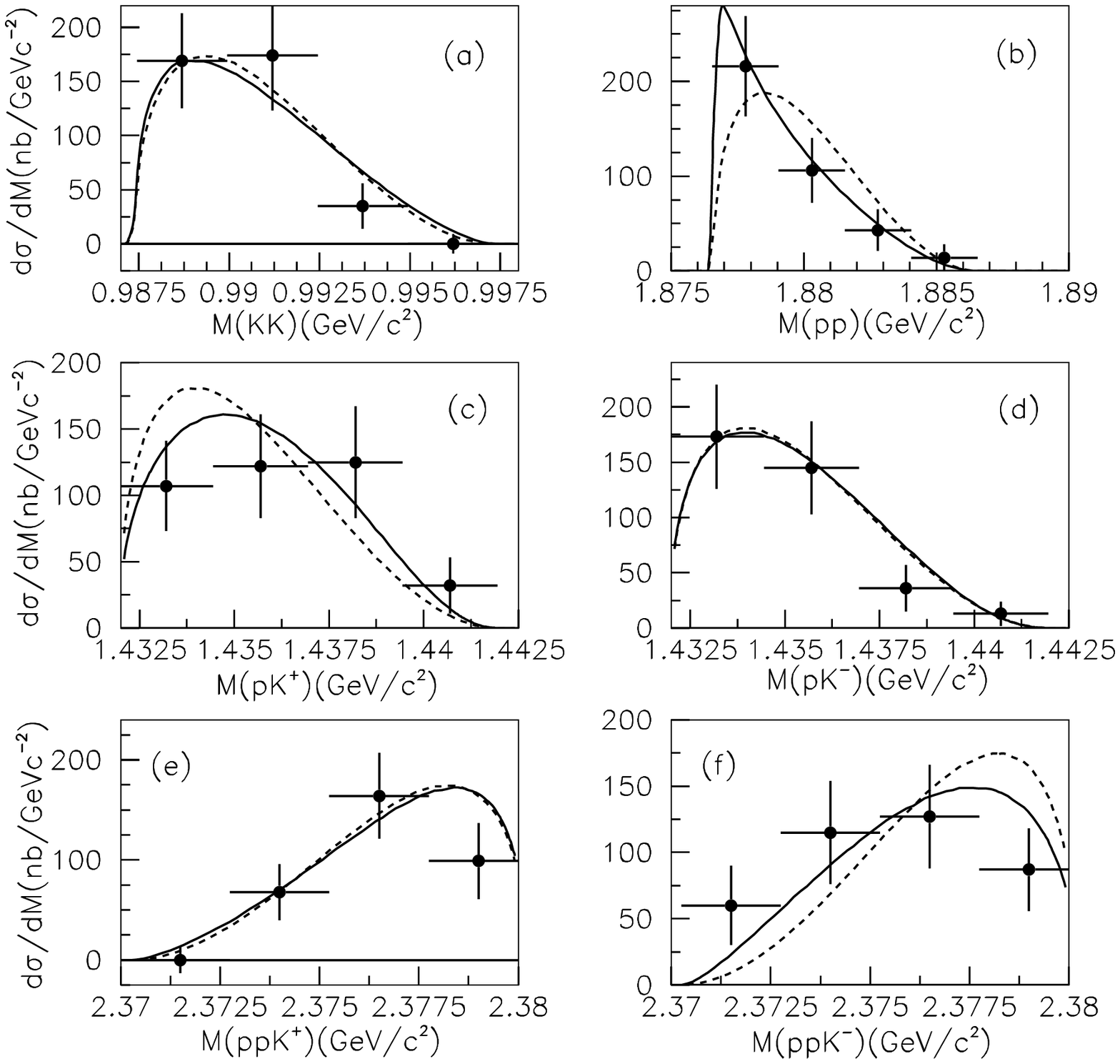}
\vspace{-0.2cm} \caption{Differential cross section for the $pp \to
pp K^+ K^-$ reaction at the excess energy of $\varepsilon$ = 10 MeV
compared with the experimental data~\cite{Silarski:2009yt}. The
solid curves stand for our theoretical calculations while the dashed
lines represent the pure phase space distribution.} \label{kk10}
\end{center}
\end{figure*}

\begin{figure*}[htbp]
\begin{center}
\includegraphics[scale=0.8]{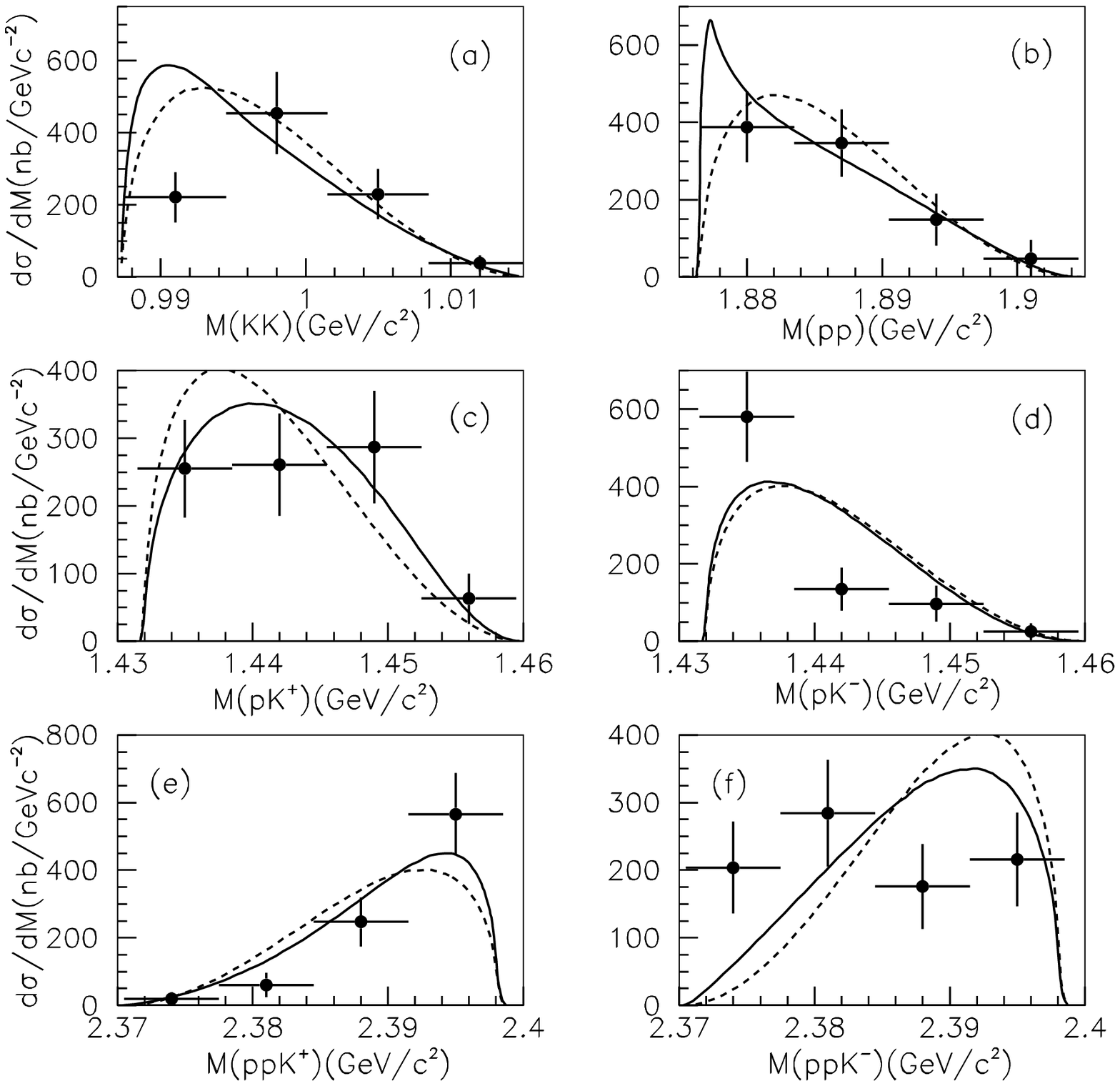}
\vspace{-0.2cm} \caption{As in Fig.~\ref{kk10} but at the excess
energy of $\varepsilon$ = 28 MeV.} \label{kk28}
\end{center}
\end{figure*}

Finally, in Figs.~\ref{kk10} and \ref{kk28}, with the parameters of
Set III, we give our model predictions of the differential cross
sections for the $pp \to ppK^+K^-$ reaction at excess energies
$\varepsilon$ = 10 and $\varepsilon$ = 28 MeV together with
experimental data~\cite{Silarski:2009yt}. It is shown that our
theoretical calculations can reasonably describe the experimental
data at both excess energies $\varepsilon$ = 10 and $\varepsilon$ =
28 MeV, especially for the $K^+K^-$ and $pp$ invariant mass
distributions, which is because we have included both the $K^+K^-$
FSI and the $pp$ FSI.

From Fig.~\ref{kk28}, one can see that, although we have considered
the contributions from the $\Lambda(1405)$ state, we still can not
well reproduce the invariant $K^- p$ mass distribution. This
indicates the strong $K^- p$ FSI. In Ref.~\cite{Geng:2007vm}, the
role of the two $\Lambda(1405)$ states which are dynamically
generated from the $\bar{K}N$ and $\pi \Sigma $ chiral
interactions~\cite{Jido:2003cb,Magas:2005vu,Oset:2001cn}, is
investigated at a proton beam p$_{\rm lab} = 3.65$ GeV
(corresponding to $\varepsilon = 108$ MeV for the case of the $pp
\to pp K^+ K^-$ reaction). It is shown that the kaon-exchange term,
which is mostly dominated by the high energy $\Lambda(1405)$ pole,
is crucial to produce the line shape of the $\pi^0 \Sigma^0$
[$\Lambda(1405) \to \pi^0 \Sigma^0$]. Thus, the kaon-exchange
mechanism should be also important in the present case, especially
for producing the line shape of the $K^- p$. However, our model can
give a reasonable description of the experimental data on the total
cross section and most differential cross sections in the considered
energy region. Meanwhile, our calculation offers some important
clues for the mechanisms of the $pp \to pp K^+K^-$ reaction and
makes a first effort to study the $K^+K^-$ FSI with the chiral
unitary approach. Hence, we will leave the issue of the strong $K^-
p$ FSI to further studies.

\section{Summary and Conclusions }

In this work, we have investigated the $pp \to ppK^+K^-$ reaction
within an effective Lagrangian model. With the assumption that the
kaon pair production is mainly through the nucleon pole,
$\Lambda(1115)$ and $\Lambda(1405)$, our calculation can reproduce
the total cross section at the energy region below the threshold of
the production of the $\phi$ meson.

We adopted the pseudoscalar and pseudovector couplings for the $\pi
NN$ interaction. It is found that both pseudoscalar and pseudovector
couplings can describe the experimental data, but, the
$\Lambda(1115)$ plays an important role in the case of the PS
coupling, while the $\Lambda(1405)$ contribution is predominant for
the PV coupling. However, considering the contributions from the
$\Lambda(1405)$ state, we still can not well reproduce the invariant
$K^- p$ mass distribution, which indicates the strong $K^- p$ FSI.

In addition to the $pp$ FSI using the Jost function, the $K^+K^-$
FSI is also studied with the chiral unitary approach, where the
scalar mesons $f_0(980)$ and $a_0(980)$ were dynamically generated
as poles of the $S$-wave amplitudes. In this sense the role of
mesons $f_0(980)$ and $a_0(980)$ are played through the $K^+ K^-$
FSI. After taking the $pp$ and $K^+K^-$ FSI into account, the
experimental data on the invariant mass distributions of the $pp$
and $K^+K^-$ are well reproduced at an excess energy $\varepsilon$ =
23.9 MeV. Besides, it can also be seen that the contribution from
the isospin zero channel is much stronger than the isospin one
channel in the $K^+K^- \to K^+K^-$ process in a chiral unitary
approach~\cite{Oller:1997ti}, which agrees with the experiment data
analysis given by Ref.~\cite{Ye:2013ndq}.

It is evident that the FSI of the four body $ppK^+K^-$ is extremely
complex. Nevertheless, taking the $pp$ and $K^+K^-$ FSI into
account, the energy dependence of the total cross sections below the
threshold of the production of the $\phi$ meson can be well
reproduced by considering the contributions from the nucleon pole,
$\Lambda(1115)$, and $\Lambda(1405)$. However, the strong $K^- p$
FSI is still required to study by further theoretical works because
it always connected with the controversial $\Lambda(1405)$ state.

\section*{Acknowledgments}

We warmly thank Qiujian Ye for sending us the experimental data
files, and Xu Cao for helpful discussions. This work is partly
supported by the National Natural Science Foundation of China under
Grant No.11105126, and the Scientific Research Foundation for the
Returned Overseas Chinese Scholars, State Education Ministry.

\bibliographystyle{unsrt}

\end{document}